\documentstyle[12pt,a4]{article}
\begin{document}

\title{Vortex line representation for flows of ideal and viscous fluids \cite{0}}
\author{E.A. Kuznetsov\footnote{ e-mail: kuznetso@itp.ac.ru}\\
{\it Landau Institute for Theoretical Physics }\\
{\it 2 Kosygin str., 119334 Moscow,  Russia }}

\date{}

\maketitle

\begin{abstract}
 It is shown that the Euler hydrodynamics for vortical  flows of an  
ideal 
 fluid coincides with the equations of motion of a  charged 
 {\it compressible} fluid moving due to a self-consistent 
electromagnetic field. 
 Transition to the Lagrangian description in a new hydrodynamics is 
equivalent 
 for the original Euler equations to the mixed Lagrangian-Eulerian 
description - 
 the vortex line representation (VLR) \cite{13}.  Due to 
compressibility of a "new" fluid
 the collapse of vortex lines can happen as the result  of breaking 
 (or overturning) of vortex lines. It is found that the Navier-Stokes 
equation 
 in the vortex line representation can be reduced to the equation of 
the diffusive type for the Cauchy invariant  with the diffusion 
tensor given by the metric of 
the VLR.     
 
\end{abstract}

\medskip


\medskip

{\bf 1.}
 Collapse as a process of a singularity formation in a finite time from 
 the initially smooth distribution plays 
 the very important role being considered as one of the most effective 
 mechanisms of the energy dissipation. 
  For hydrodynamics of incompressible fluids collapse must play 
  also a very essential role.
It is well known that appearance of singularity in gasodynamics, i.e., 
in compressible hydrodynamics, 
is connected with the phenomenon of breaking  that is the physical mechanism leading to 
emergence  of shocks. From the point of 
view of the classical catastrophe theory 
\cite{arnold}  this process is nothing more than the formation of folds.
It is completely characterized by the mapping corresponding to 
transition from the 
Eulerian description to the Lagrangian one.   Vanishing the Jacobian $J$ of 
this mapping means emergence 
of a singularity for spatial derivatives of velocity and density of a gas.  
In the incompressible case
breaking as intersection of trajectories of Lagrangian particles 
is absent because the Jacobian of the corresponding mapping is fixed, 
in the simplest case
equal to unity. By this reason, it would seem that 
there were no any reasons for existence of such phenomenon at all. 
In spite of this fact, 
as it was shown in \cite{10,11,12}, breaking,  however,  
is possible in this case also.  
It can happen with vortex lines. Unlike  the  breaking in gasodynamics,  the breaking 
of vortex lines means that
one vortex line reaches another vortex line.  For smooth initial  conditions
breaking happens first time while touching vortex lines at  a single  point. 
In the touching point
 the vorticity becomes infinite. And this is possible in spite of 
incompressibility of both divergence-free fields, i.e., vorticity and velocity.  
To describe the 
breaking of vortex lines in the papers \cite{13,14} it was suggested 
the vortex line representation 
 -- a mixed Lagrangian-Eulerian description  when each vortex line is 
labeled by a two-dimensional marker and another parameter defines 
the vortex line itself. 

This paper is devoted to development of this method to  apply to both
ideal and viscous fluids. We clarify the role of the Clebsch variables 
in the vortex line representation: these variables can be used as Lagrangian markers 
of vortex lines. However, as well known, 
these variables can be introduced always only locally 
and, generally speaking, can not be extended for the whole space. 
In the general situation  we demonstrate in this paper
that transition to the vortex line representation is equivalent to consideration of 
a new {\it compressible} hydrodynamics of a charged fluid flowing under action of 
a self-consistent electromagnetic field. In this case the electric and magnetic 
fields satisfy the Maxwell equations. The most essential property of a new hydrodynamics 
is a compressibility of a new fluid that for its Lagrangian 
description means compressibility of
the corresponding mapping and, respectively, a possibility of 
a breaking. In terms of the Eulerian 
characteristics this results in  the  breaking of vortex lines 
when the vorticity 
${\bf \Omega}= \mbox{curl}~{\bf v}$ takes  infinite value.  
In the framework of the new hydrodynamics
 of a  charged fluid  the role of density plays the quantity 
inverse to $J$  which is naturally called
as a density of vortex lines.  This quantity appears from the 
Cauchy formula for the vorticity
${\bf \Omega}$. Evolution of the vortex line 
density in time and space  is defined by the velocity 
component normal to the vorticity. As it is shown in this paper  
the Cauchy formula can be 
obtained from a ``new'' Kelvin theorem as well as from the analog 
of the Weber transformation. As the result, the Euler equations turn out 
to be resolved with respect to the Cauchy invariants, i.e., relative to 
the infinite number of integrals of motion.  In this case one can consider 
the Euler equations as the partially integrated equations.  
This circumstance  is very important 
for numerical solution of the Euler equation. 

The vortex line representation can be applied not only to ideal 
hydrodynamics  but also to  flow description 
of viscous incompressible fluids in the framework of the Navier-Stokes equation. 
In the paper we obtain the 
equation of the diffusion type  describing dynamics of the Cauchy invariant in 
the viscous case with 
the ``diffusion tensor'' determined by the VLR metric. In  its form this 
equation coincides with 
the equation derived in \cite{Z}.  In this case the equations of motion 
of vortex lines  in its 
original (for ideal fluids) form are understood as the equations 
given the transformation to 
a new curvilinear 
system of coordinates.  The obtained exact equations for description 
of viscous flows can be considered
as the result of exact separation of two different temporal scales:
the inertial (in fact, nonlinear) scale  and the viscous one.

\vspace{0.5cm}

{\bf 2.}  As  well known ( see, for instance, \cite{16}, \cite{17}) the Euler equations 
for an ideal incompressible fluid,  
\begin{equation}
\label{euler}
\frac{\partial {\bf v}}{\partial t}+
({\bf v}\nabla ){\bf v}=
-\nabla p, \qquad
\mbox {\rm div}~{\bf v}=0,
\end{equation}
in both two-dimensional and three-dimensional cases possess
the infinite (continuous) number of integrals of motion.  These are 
the so called  Cauchy invariants. The most simple way to derive the Cauchy invariants 
is one to use
the Kelvin theorem about conservation of the velocity circulation, 
\begin{equation}
\label{9}
\Gamma = \oint ({\bf v}\cdot d{\bf l}), 
\end{equation}
where the integration contour $C[{\bf r}(t)]$  moves together with a fluid.
If in this expression one makes a  transform from the Eulerian coordinate  ${\bf r}$  to 
the Lagrangian ones  ${\bf a}$ then Eq.  (\ref{9})
can be rewritten as follows: 
\begin{displaymath}
\label{9a}
\Gamma =\oint \dot x_i\cdot \frac{\partial x_i}{\partial a_k}~
da_k ~,
\end{displaymath}
where a new contour $C[{\bf a}]$ is already immovable.
Hence,  due to arbitrariness of the contour $C[{\bf a}]$ and using the 
Stokes formula 
one can conclude that the quantity 
\begin{equation}
\label{10}
{\bf I} = \mbox {\rm rot}_a~\Biggl ( \dot x_i
\frac{\partial x_i}{\partial {\bf a}} \Biggr )
\end{equation}
conserves in time at each point  ${\bf a}$.  This is just 
the Cauchy invariant.
If the Lagrangian coordinates ${\bf a}$ in (\ref{10}) coincide with the 
initial positions
of fluid particles  the invariant  ${\bf I}$ is equal to the initial vorticity  
${\bf \Omega}_0({\bf a})$.

Conservation of these invariants, as it was shown first  by 
Salmon  \cite{17},  is consequence of the special (infinite) 
symmetry - the so-called 
relabeling symmetry.   The Cauchy invariants characterize the frozenness 
of the vorticity into fluid.
 This is a very important property according to which fluid 
(Lagrangian) particles
can not leave its own vortex line where they were initially. Thus, the 
Lagrangian particles  
have one independent degree of freedom --  motion along vortex line.  
From another side, 
such a motion as it follows from the equation  for the vorticity
\begin{equation}
\label{8}
\frac{\partial{\bf \Omega}}{\partial t}=
\mbox {\rm rot}~ [{\bf v}\times{\bf \Omega} ], 
\end{equation}
does not change its value. From this point of view a vortex line represents
the invariant object and therefore it is natural to seek for such 
a transformation 
when this invariance is seen from the very beginning.
Such type of description - the vortex line representation -  was introduced 
in the papers \cite{13,14}
by  Ruban and the author of this paper.

\vspace{0.5cm}

{\bf 3.}
 Consider the vortical flow  (${\bf \Omega}\neq 0$) of an ideal 
fluid given 
by the Clebsch variables 
$\lambda$ and  $\mu$:
\begin{equation}
\label{15}
{\bf \Omega}=
[\nabla\lambda\times\nabla\mu ].
\end{equation} 
The geometrical meaning of these variables is well known: intersection 
of two surfaces 
$\lambda = \mbox {\rm const}$ and $\mu = \mbox {\rm const}$
yields the vortex line. It is known also that in the incompressible case 
the Clebsch variables 
are Lagrangian invariants, being unchanged along trajectories of 
fluid particles:
\begin{equation}
\label{16}
\frac{\partial\lambda}{\partial t}+
({\bf v}\nabla )\lambda =0; \quad\quad
\frac{\partial\mu}{\partial t}+
({\bf v}\nabla )\mu =0.
\end{equation}
Therefore these variables can be taken as markers for vortex lines.
It is easily to establish that transition in 
(\ref{15}) to new variables 
\begin{equation}
\label{17}
\lambda = \lambda (x,y,z), \qquad
\mu = \mu (x,y,z), \qquad
s=s (x,y,z),
\end{equation}
where $s$  is the parameter given the vortex line, leads  to  the expression
\begin{equation}
\label{13}
{\bf \Omega}({\bf r},t)= \frac 1J \cdot
\frac{\partial {\bf R}}{\partial s}
\end{equation}
where
\begin{equation}
\label{14}
J=\frac{\partial(x,y,z)}
{\partial(\lambda,\mu, s)} 
\end{equation}
is the Jacobian of the mapping
\begin{equation}
\label{11}
{\bf r}={\bf R}(\lambda,\mu,s).
\end{equation}
The transform (\ref{11}) inverse to (\ref{17})  defines the corresponding transition 
to the curvilinear, connected with vortex lines,
system of coordinates. 

The equations of motion of vortex lines - the equations for 
${\bf R}(\lambda,\mu,s,t)$ -- 
can be obtained directly from the equation of motion for the vorticity
(\ref{8}).  However, the most simple way to derive them is to use the combination 
of the equations  
(\ref{16}):
\begin{equation}
\label{20}
\nabla\mu \Biggl [ \frac{\partial\lambda}{\partial t}+
({\bf v}\nabla )\lambda \Biggr ] -
\nabla\lambda \Biggl [ \frac{\partial\mu}{\partial t}+
({\bf v}\nabla )\mu \Biggr ] =0,
\end{equation}
which is identical to (\ref{16}) due to a linear 
independence of the vectors
$\nabla\lambda$ and $\nabla\mu$.

 Performing in  (\ref{20}) the transformations (\ref{17}), we arrive 
at the equation of motion for vortex lines \cite{13}:
\begin{equation}
\label{21}
\Biggl [ \frac{\partial {\bf R}}{\partial s}\times
\Biggl ( \frac{\partial {\bf R}}{\partial t}-
{\bf v}({\bf R},t) \Biggr ) \Biggr ] =0.
\end{equation}
 This equation has one important property: any motion along 
a vortex line does not change 
the line itself. It is easily to check that Eq. (\ref{21}) is 
equivalent to the equation
\begin{equation}
\label{mapping}
\frac{\partial {\bf R}}{\partial t}=
{\bf v}_n({\bf R},t), 
\end{equation}
 where ${\bf v}_n$ is the velocity component normal to the 
vorticity vector.

In accordance with the Darboux theorem,  the Clebsch variables 
can be introduced 
locally always but not globally. 
It is well known also that the flows parameterized by the Clebsch 
variables has a zero helicity integral
$\int ~({\bf v}\cdot\mbox {\rm rot}~{\bf v}) d{\bf r}$ -- the 
topological invariant  
which characterizes a degree of knottiness of vortex lines. 
Therefore to introduce the vortex line representation for flows 
with nontrivial topology
it is necessary to come back  to the original equations of motion 
(\ref{euler}) and (\ref{8}) for velocity and vorticity.

\vspace{0.5cm}

{\bf 4.}  According to the equation (\ref{8}) the tangent 
to the vector ${\bf \Omega}$ 
velocity 
component ${\bf v}_{\tau}$ does not effect (directly) on the 
vorticity dynamics, i.e., in
(\ref{8}) we can put, instead of  
${\bf v}$, its transverse component ${\bf v}_n$.

The equation of motion for the transverse velocity ${\bf v}_n$  
follows directly
from the equation 
(\ref{euler}). It has the form of the equation of motion of 
charged particle
moving in an electromagnetic field:
\begin{equation}
\label{electron}
\frac{\partial {\bf v}_n}{\partial t}+
({\bf v}_n\nabla){\bf v}_n=
{\bf E}+[{\bf v}_n\times {\bf H}],
\end{equation}
where the effective electric and magnetic fields are given by 
the expressions:
\begin{equation}
\label{electric}
{\bf E}=-\nabla \left ( p+\frac{v^2_{\tau}}{2} \right )-
\frac{\partial {\bf v}_{\tau}}{\partial t},
\end{equation}
\begin{equation}
\label{magnetic}
{\bf H}=\mbox {\rm rot}~{\bf v}_{\tau}.
\end{equation}
Interesting to note that the electric and magnetic fields 
introduced above
are expressed through the scalar $\varphi$ and vector 
${\bf A}$ potentials
 by the standard way: 
\begin{equation}
\label{potentials}
\varphi = p+\frac{{\bf v}^2_{\tau}}{2}, \qquad
{\bf A}={\bf v}_{\tau},
\end{equation}
 so that two Maxwell equations 
\begin{displaymath}
\mbox {\rm div}~{\bf H}=0, \qquad
\frac{\partial {\bf H}}{\partial t}=
-\mbox {\rm rot}~{\bf E}
\end{displaymath}
satisfy automatically.
In this case the vector potential ${\bf A}$ 
has the gauge 
$$
\mbox{div}~ {\bf A}=- \mbox{div}~{\bf v}_n,
$$
which is equivalent to the condition $\mbox{div}~{\bf v}=0$.

Two other Maxwell equations can be written also but they  
can be considered as
definition of the charge density $\rho$ and the current  
${\bf j}$ which follow from 
the relations  
(\ref{electric}) and (\ref{magnetic}).
 The basic equation in the new hydrodynamics is the  equation 
of motion  (\ref{electron})
for the normal component of the velocity which represents 
the equation of motion for nonrelativistic particle with 
a charge and a mass
equal to unity, the light velocity in this units is equal to $1$.

The equation of motion (\ref{electron}) is written in the 
Eulerian representation. 
To transfer to its Lagrangian 
formulation one needs to consider the equations for 
"trajectories" given by the 
velocity ${\bf v}_n$:  
\begin{equation}
\label{mapping1}
\frac{d {\bf R}}{d t}={\bf v}_n ({\bf R},t)
\end{equation}
with initial conditions
$$
{\bf R}|_{t=0}={\bf a}.
$$ 
Solution of the equation (\ref{mapping1}) yields the mapping
\begin{equation}
\label{19}
{\bf r}={\bf R}({\bf a},t),
\end{equation}
which defines transition from the Eulerian description 
to a new Lagrangian one.

 The equations of motion in new variables are the Hamilton equations:
\begin{equation}
\label{ham}
\dot{\bf P}=-\frac{\partial h}{\partial {\bf R}}, \qquad
\dot{\bf R}=\frac{\partial h}{\partial {\bf P}},
\end{equation}
where dot means differentiation with respect to time for fixed ${\bf a}$, 
${\bf P}={\bf v}_n+{\bf A}\equiv {\bf v}$ is the generalized momentum, and the 
Hamiltonian of a particle
$h$ being a function of momentum  ${\bf P}$ and coordinate ${\bf R}$ is given 
by the standard expression:
\begin{displaymath}
h= \frac 12 ({\bf P}- {\bf A})^2 + \varphi \equiv p+\frac{{\bf v}^2}{2},
\end{displaymath}
i.e.,  coincides with the Bernoulli "invariant".

 The first equation of the system  (\ref{ham}) is the equation 
 of motion  (\ref{electron}),
written in terms of ${\bf a}$ and $t$, and the second equation 
coincides with  (\ref{mapping1}). 

 For new hydrodynamics (\ref{electron}) or for its Hamilton version (\ref{ham})
 it is possible to formulate a "new" Kelvin theorem (it is also the Liouville theorem):
\begin{equation}
\label{kelvin}
\Gamma = \oint ({\bf P }\cdot d{\bf R }), 
\end{equation}
where integration is taken along a loop moving 
together with the "fluid". Hence, analogously as it was made before while derivation of
(\ref{10}) we get the expression for a new Cauchy invariant:
\begin{equation}
\label{18}
{\bf I} = \mbox {\rm rot}_a~\Biggl ( P_i
\frac{\partial x_i}{\partial {\bf a}} \Biggr ).
\end{equation}
 Its difference from the original Cauchy invariant
 (\ref{10})  consists in that in the equation of motion  (\ref{mapping1})
 instead of the velocity ${\bf v}$ stands
its normal component ${\bf v}_n$. As consequence, the  "new" hydrodynamics 
becomes compressible:
 $\mbox {\rm div}~{\bf v}_n \neq 0$. Therefore on the Jacobian  $J$ of the mapping
  (\ref{19}) there are imposed no restrictions. The Jacobian $J$ can take arbitrary values.

From the formula  (\ref{18}) it is easily to get the expression for the 
vorticity ${\bf \Omega}$ in the given point
${\bf r}$ at the instant $t$   (compare with \cite{13,14}):
\begin{equation}
\label{cauchi}
{\bf \Omega}({\bf r},t)=
\frac{({\bf \Omega}({\bf a})\cdot\nabla_a)
{\bf R}(a,t)}{J},
\end{equation}
where $J$ is the Jacobian of the mapping (\ref{19}) equal to 
\begin{displaymath}
J=\frac{\partial(x_1,x_2,x_3)}{\partial(a_1,a_2,a_3)}.
\end{displaymath}
Here we took into account that the generalized momentum  ${\bf P}$ coincides with
the velocity 
${\bf v}$, including the moment of time  $t=0$:
${\bf P}_0({\bf a})\equiv {\bf v}_0({\bf a})$.  ${\bf \Omega}_0({\bf a})$
 in this relation is the "new" Cauchy invariant
with zero divergence:
$\mbox~{\rm div}_a {\bf \Omega}_0(a)=0$.

 The representation  (\ref{cauchi}) generalizes the relation (\ref{15}) 
 to an arbitrary  topology of  vortex lines.  The variables ${\bf a}$  
in this expression can be considered 
 locally  as a set of 
 $\lambda$, $\mu$ and $s$. 

As known  (see, for instance, \cite{14}), expression for the Cauchy invariant
 can be obtained from the Weber transformation. This is the representation of velocity
 in terms of the initial data which can be obtained by integrating 
 the Cauchy invariant (\ref{cauchi}). 

Consider the following one-form  $\omega=({\bf P }\cdot d{\bf R })$ and calculate 
its time derivative. 
By means of the equations of motion  (\ref{ham}) we get:
$$
\dot\omega= d[-h+({\bf P }\dot{\bf R })].
$$
Hence it follows that the vector function
$$
u_k=\frac{\partial x_i}{\partial a_k}\cdot P_i,
$$
dependent on $t$ and  ${\bf a}$, will obey the following equation of motion:
\begin{displaymath}
\dot u_k=\frac{\partial}{\partial a_k}
\left ( -p+\frac {v^2_n}{2}- \frac {v^2_{\tau}}{2}\right ).
\end{displaymath}
Integration of this equation in time gives the Weber-type transformation:
\begin{equation}
\label{weber}
u_k({\bf a},t)=
u_{k0}({\bf a})+
\frac{\partial\Phi}{\partial a_k},
\end{equation}
where the potential  $\Phi$ satisfies the nonstationary Bernoulli equation:
$$
\dot\Phi = -p+\frac {v^2_n}{2}- \frac {v^2_{\tau}}{2}.
$$
If $\Phi |_{t=0}=0$ then the time independent vector 
${\bf u}_0({\bf a})$ coincides with the initial velocity
${\bf v}_0({\bf a})$. By applying the operator curl to the relation  (\ref{weber})
we arrive again at the Cauchy invariant (\ref{18}).
 
Thus, in the general situation the equation of motion of vortex lines has the form 
(\ref{mapping1}) which is completed by the relation  
(\ref{cauchi}) and the equation 
\begin{equation}
\label{24}
{\bf \Omega}({\bf r},t)=
\mbox {\rm rot}_r{\bf v}({\bf r},t)
\end{equation}
with additional constraint $\mbox {\rm div}_r{\bf v}({\bf r},t)=0$.

 The equations of motion  (\ref{mapping1}), (\ref{24}) 
together with the relation  (\ref{cauchi})  
can be considered as the result of partial integration of the Euler equation 
(\ref{euler}).  These new equations are resolved with respect to the 
Cauchy invariants -- an infinite number of integrals of motion, that is a very 
important issue for numerical integration
(see \cite{11,12}).  For the  partially integrated system 
the Cauchy invariants conserve automatically that, however,
for direct numerical integration of the Euler equation  one needs
 to test in which extent these invariants 
remain constant. Probably, this is one of the main restrictions defining 
accuracy of discrete algorithms of direct integration of the
Euler equations.

Another very important property of the vortex line representation is absence 
of any restrictions on the value of the Jacobian $J$ which do exist, for instance, 
for transition from the Eulerian description to
the Lagrangian one in the original Euler equation 
 (\ref{euler}) when Jacobian in the simplest situation is equal to unity. 
The value $1/J$ for the system (\ref{mapping1}), (\ref{24}), (\ref{cauchi}) 
has a meaning of a density $n$ of vortex lines.
This quantity as a function of ${\bf r}$ and $t$ , according to 
 (\ref{mapping1},  obeys the discontinuity equation:
\begin{equation}
\label{den}
 \frac{\partial n}{\partial t}+\mbox {\rm div}_r(n{\bf v}_n) =0.
\end{equation}
In this equation  $\mbox {\rm div}_r{\bf v}_n\neq 0$ because only 
the total velocity has zero divergence.

{\bf 5.}  Consider now the question about application of the VLR to flows 
of viscous fluids. 
Write down the Navier-Stokes equation for  vorticity  ${\bf \Omega}$:
\begin{equation}
\label{NS}
\frac{\partial {\bf \Omega}}{\partial t}=
\mbox {\rm rot} [{\bf v}\times {\bf \Omega}]-
\nu~ \mbox  {\rm rot}~\mbox {\rm rot}~{\bf \Omega},
 \end{equation}
and perform in this equation the transform to new variables 
 ${\bf a}$ and $t$ by means of changes defined by 
the equation  (\ref{mapping1}) together with the Cauchy relation
 (\ref{cauchi}) where 
${\bf\Omega}_0$ is assumed  a function of not only ${\bf a}$ but 
also time $t$: ${\bf\Omega}_0=
{\bf\Omega}_0({\bf a},t)$.

Then after substitution  (\ref{cauchi}) into  (\ref{NS}) the 
first term in the right hand side is cancelled
 because of (\ref{mapping1}). At the result,  the equation 
(\ref{NS}) is written in the form:
\begin{equation}
\label{NS2}
\frac 1J \left ( \frac{\partial{\bf \Omega}_0}
{\partial t} \cdot \nabla_a \right ) {\bf R}=
-\nu ~\mbox  {\rm rot}~\mbox {\rm rot}\left \{\frac 1J
({\bf \Omega}_0\cdot\nabla_a ){\bf R}\right\}.
\end{equation}

Next, change differentiation relative to ${\bf r}$  in 
the r.h.s. of  (\ref{NS2})  to differentiation
against  ${\bf a}$.  After simple, but cumbersome calculations 
the equation (\ref{NS2}) transforms into 
the equation for ${\bf\Omega}_0({\bf a},t)$:
\begin{equation}
\label{NS3}
\frac{\partial{\bf \Omega}_0}
{\partial t}=-\nu~\mbox {\rm rot}_a
\left (\frac {\hat g}{J}\mbox{\rm rot}_a \left (\frac {\hat g}{J}
{\bf \Omega_0}\right)\right).
\end{equation}
Formally it is a linear equation for  ${\bf\Omega}_0$,
here  $\hat g$  is the metric tensor equal to
\begin{displaymath}
g_{\alpha\beta}=\frac{\partial x_i}{\partial a_{\alpha}}\cdot
\frac{\partial x_i}{\partial a_{\beta}}.
\end{displaymath}

 The equation (\ref{NS3}) for the Cauchy invariant formaly 
 coincides with that obtained by Zenkovich and Yakubovich for incompressible
 hydrodynamics  \cite{Z} in which  the variables ${\bf a}$  are assumed to 
 be Lagrangian markers
 of fluid particles.  In the Zenkovich-Yakubovich equation the Jacobian $J$ 
 is proposed to be independent
 on time, in the simplest case equal to 1. Just this is a principle difference between 
the Zenkovich-Yakubovich equation and the equation (\ref{NS3}). $J$ 
in (\ref{NS3}) is a function of time $t$ and coordinates ${\bf a}$.

 Remarkable peculiarity of the obtained system is the {\it exact separation}
of two  different temporal scales, responsible for the inertial (in fact, nonlinear) 
processes and for the viscous processes. The former ones  are described by 
the equation  (\ref{mapping1}), and the latter by the equation of diffusive type 
(\ref{NS3}) in which the diffusion "coefficient", proportional to viscosity $\nu$,      
is defined by the metric of the mapping 
${\bf r}={\bf R}({\bf a},t)$.

\vspace{0.5cm}

The author is grateful to E.I. Yakubovich for possibility of acquaintance 
with the paper  \cite{Z} before its publication.
This work was supported by the RFBR (grant no. 00-01-00929),
by the Program of Support of the Leading Scientific Schools of Russia
(grant no.~00-15-96007) and by INTAS (grant no. 00-00797).

\end{document}